\documentclass[12pt]{iopart}

\usepackage{amssymb}

\newcommand{\bb}{\bibitem}
\newcommand{\be}{\begin{eqnarray}}
\newcommand{\ee}{\end{eqnarray}}
\newcommand{\rar}{\rightarrow}

\begin{document}

\title{Brane Cosmology and KK Gravitinos}

\author{C.~Bambi, F.R.~Urban}
\address{University of Ferrara, Department of Physics \\
INFN, sezione di Ferrara \\
via Saragat 1, 44100, Ferrara, Italy}

\date{\today}

\begin{abstract}
The cosmology of KK gravitinos in models with extra dimensions is considered. The main result is that the production of such KK modes is not compatible with an epoch of non--standard expansion after inflation. This is so because the BBN constraint on the zero mode forces the reduced five dimensional Planck mass $M_5$ down to values much smaller than the usual four dimensional one, but this in turn implies many KK states available for a given temperature. Once these states are taken into account one finds that there is no $M_5$ for which the produced KK gravitinos satisfy BBN and overclosure constraints. This conclusion holds for both flat and warped models in which only gravity propagates in the full spacetime.
\end{abstract}

\section{Introduction}\label{intro}

The problem of overproduction of gravitinos, the supersymmetric partner of the graviton, is a long--standing one in cosmology \cite{problem}. The gravitino interacts very weakly with ordinary matter, its coupling being gravitationally suppressed, and this makes it a long living particle which is never in thermal equilibrium after inflation. 

An unstable gravitino lighter than about 10 TeV decays after the Big Bang Nucleosynthesis (BBN) and the entropy injected into the plasma can cause photodissociation of the light elements, altering their abundances \cite{abundance}. Hence the requirement of a successful BBN severely constrains the produced amount of gravitinos \cite{abundance}.

If gravitinos are much heavier, their decay products are not dangerous for primordial nuclei since they harmlessly decay before BBN; however, if $\cal R$--parity is a good symmetry (as it is assumed throughout the whole paper) their decay will produce a non--thermal abundance of light SUSY particles, either the lightest stable one (LSP), or some other particle which will later decay into it. The present day energy density stored in LSP as dark matter is constrained by cosmological observations \cite{wmap}. A similar scenario holds for a light gravitino, lighter than about 100 GeV, which, if it is the LSP as it is likely is the case, must (at least) not overclose the universe. These considerations lead to an upper limit on the temperature at which thermal equilibrium had been established (usually referred to as the reheating temperature $T_R$), this limit being around $10^3\div10^6$ TeV, depending on the model \cite{abundance}. This is discussed in the forthcoming section.

Of course this picture is drastically modified in supersymmetric extra dimensional models \cite{xsusyA, xsusyB, xsusyC}, in which case one or more KK towers of gravitinos have to be taken into account: it is expected that these extra states will more seriously constrain the allowed maximum temperature reached in the early universe. Furthermore, in braneworlds, where all of the Minimal Supersymmetric Standard Model (MSSM) particles are forced to live on a brane, while (super)gravity is effectively five dimensional, the Friedman equation contains extra terms \cite{b-cosmology}, which modify the standard cosmological expansion and consequently the picture of gravitino production. In this paper the cosmology of such KK states is studied, in as general as possible frameworks, during both standard and non--standard expansion regimes.

Some details about the gravitino problem are presented in section \ref{prob}, including its ``standard'' solution (sec.\ \ref{stdsol}) and the ``braneworld'' one (sec.\ \ref{branesol}), first proposed in \cite{osamu}. Then, a brief review on SUSY extra dimensional models is given, focussing on the features of the models to be dealt with (section \ref{xsusy}). With the setup at hand, the bounds on the five dimensional Planck mass for both flat and warped models are calculated, first for an epoch of standard expansion (sec.\ \ref{stdkk}), and then in the alternative case (sec.\ \ref{branekk}). Other possible production mechanisms are considered in section \ref{mech}, while in section \ref{gravitons} constraints arising from KK gravitons decay are discussed. The last section is devoted to a short summary and contains the conclusions.

\section{The gravitino problem}\label{prob}

Gravitinos are produced in several different ways, thermally and non--thermally. Thermal production \cite{thermal} involves either inelastic scattering processes of thermalised particles, or decays of supersymmetric particles. For the reason explained in section \ref{decay}, this last mechanism is uninteresting in braneworld cosmology. Non--thermal mechanisms \cite{non-thermal} include perturbative and non--perturbative production by means of inflaton decay, or some other scalar fields (moduli, dilaton, radion), which is strongly model dependent, and which is not treated here, see section \ref{nonth}. The last option is gravitational particle production, which is discussed in \ref{grav}. Hence, in what follows mainly thermal production of gravitinos via inelastic scattering is considered.

The zero mode gravitino abundance is usually expressed in terms of the gravitino number density to the entropy density ratio as
\be\label{Ydef}
Y^0_{3/2}(T) = \frac{n^0_{3/2}(T)}{s(T)}\, .
\ee
Here $s(T) = (2\pi^2/45) g_{*S} T^3$ and $g_{*S}$ is the number of ``entropic'' degrees of freedom, and $T$ is the temperature of the system.

The Boltzmann equation for the process under examination leads to the abundance of the produced particles
\be\label{boltzdef}
\frac{d}{dT} Y^0_{3/2} = - \frac{s \langle \sigma v \rangle\, Y_{rad}^2}{H T}\, ,
\ee
where $Y_{rad}$ is the equilibrium number density to entropy density ratio for relativistic particles, and $\langle \sigma v \rangle$ parametrises the thermally averaged cross section for the process under scrutiny. $H$ is the Hubble parameter which in standard cosmology is $H = (\rho/3M_4^2)^{1/2}$, where $\rho$ is the energy density of the universe and $M_4=2.4\cdot10^{15}{\rm TeV}$ is the reduced four dimensional Planck mass. In the radiation dominated epoch of the early universe $\rho \propto T^4$.

Integrating this equation one finds the well known \cite{abundance} expression for the abundance at the BBN (given that the zero mode is not too heavy and thus survives at least till $T\simeq1$ MeV)
\be\label{zeroYstd}
Y_{3/2}^0 = 1.9 \cdot 10^{-19}\, \left( 1 + \frac{\tilde m^2}{3 {m_{3/2}^0}^2} \right)\, \left( \frac{T_R}{\rm TeV} \right)\, ,
\ee
where $\tilde m$ is the gluino mass.

\subsection{The standard cosmological solution}\label{stdsol}

Since the primordial gravitino abundance (\ref{zeroYstd}) is proportional to the reheating temperature, cosmological constraints on $Y_{3/2}^0$ translate into upper bounds on $T_R$ and hence on the inflationary model. If the gravitino is stable, its energy density today must not overclose the universe. In particular, it must not exceed the dark matter energy density. This puts a bound on $T_R$ only for $m_{3/2} \gtrsim 1$ keV. On the other hand, if the gravitino is unstable its decay products can alter BBN predictions and/or the CMBR spectrum. The resulting constraint depends on several unknown parameters, such as gravitino lifetime and branching ratio. If the main decay mode is hadronic, i.e.\ $B_h \approx 1$, BBN can provide very tight bounds, especially for $m_{3/2} \sim 1$ TeV, see \cite{abundance}. However, even if the main decay mode is photon + neutralino, because gravitino is lighter than the lightest color superparticle, $B_h$ is non--vanishing: the photon can be virtual and can decay into $q{\bar q}$ pair, with branching ratio $B_h \sim \alpha_{em}/4\pi \sim 0.001$. Bounds on gravitino abundances for $m_{3/2} = 100 \; {\rm GeV} \div 30 \; {\rm TeV}$ are summarised in the table below for $B_h = 1$ [$B_h = 10^{-3}$[. More details are given in \cite{abundance}.

\begin{center}
\begin{tabular}{|c c|c|c|}
\hline
$m_{3/2}$ & $Y_{3/2}$ &	$m_{LSP}$ & constraint \\
\hline
$1 \; {\rm keV} \div 100 \; {\rm GeV}$ & 
$6\cdot10^{-10} \left(\frac{1 \; {\rm GeV}}{m_{3/2}}\right)$ & 
$m_{3/2}$ & direct overclosure \\
$100 \; {\rm GeV}$ & $3\cdot10^{-16} \; [3\cdot10^{-16}]$ & 
$\;$ any but $m_{3/2} \;$ & BBN \\
$300 \; {\rm GeV}$ & $4\cdot10^{-16} \; [4\cdot10^{-16}]$ & 
any but $m_{3/2}$ & BBN \\
$1 \; {\rm TeV}$ & $4\cdot10^{-17} \; [3\cdot10^{-14}]$ &
any but $m_{3/2}$ & BBN \\
$3 \; {\rm TeV}$ & $1\cdot10^{-16} \; [3\cdot10^{-14}]$ &
any but $m_{3/2}$ & BBN \\
$10 \; {\rm TeV}$ & $5\cdot10^{-15} \; [2\cdot10^{-13}]$ & 
any but $m_{3/2}$ & BBN \\
$30 \; {\rm TeV}$ & $8\cdot10^{-15} \; [2\cdot10^{-12}]$ & 
any but $m_{3/2}$ &	BBN \\
$30 \; {\rm TeV} \div 10^5 \; {\rm TeV}$ & 
$6\cdot10^{-12} \left(\frac{100 \; {\rm GeV}}{m_{LSP}}\right)$ & 
any but $m_{3/2}$ & daughters overclosure \\
\hline
\end{tabular}
\end{center}

Gravitino decay can also affect the CMBR spectrum if gravitino lifetime is in the range $10^6 - 10^{13}$ s, causing Bose--Einstein and/or Compton distortion, but the constraints are usually weaker than the BBN ones \cite{cmbr}. Nevertheless, CMBR bounds could be relevant for the case of unstable and very light gravitinos. This might be possible in extra dimensions if the LSP is an ultra--light zero mode gravitino and there are KK gravitinos with masses a little heavier. However this scenario is probably not very realistic. For $m_{3/2} \gtrsim 30$ TeV, the gravitino lifetime is $\tau_{3/2} \lesssim 1$ s and no relevant limits are obtained from BBN. If the latter is the case, the strongest bound is derived by requiring that the LSP, the stable relic produced eventually at the end of the gravitino decay chain, does not (at least) overclose the universe: in this case the result is independent of gravitino properties and is set only by LSP mass, which should be reasonably close to 100 GeV.

For instance, in the typical case where the gravitino is not the LSP and $m_{3/2} \sim 100 \; {\rm GeV} \div 1 \; {\rm TeV}$, the reheating temperature must be
\be\label{example-T}
T_R \lesssim 10^5 - 10^8 \; {\rm GeV}
\ee
and several inflationary models have to be rejected or strongly fine--tuned.

\subsection{The braneworld cosmological solution}\label{branesol}

So far only the standard cosmological expansion law $H^2=\rho/3M_4^2$ has been considered, and it has been shown that the limits on the reheating temperature are very restrictive, especially in connection with inflation model building (it is non--trivial to lower the reheating temperature down to such values and still have successful inflation\footnote{At the moment good candidates in this direction are low scale gravity models, which may ``naturally'' provide reasonably low temperatures \cite{newton}.}).

However, if the SM lived on a four dimensional Friedman--Robertson--Walker hypersurface (the brane), embedded in an extra dimensional spacetime, the early universe would admit an epoch of non--standard expansion \cite{b-cosmology}. Several such models have been built in the last few years, the ones which are dealt with here being the ADD \cite{add} and RS \cite{rs} models, which involve flat and warped extra dimensions respectively. These models show a peculiar feature when their cosmology is investigated. For instance the model named ``RSII'', where only one four dimensional three--brane is present to catalyse gravity and warp the bulk five dimensional $AdS_5$ spacetime, has a Friedman equation of this kind\footnote{This form for the Friedman equation holds more generally for five dimensional brane models, the main differences being extra terms such as cosmological constants, dark radiation, etc., which are neglected here.} \cite{b-cosmology}:
\be\label{modFried}
H^2 = \frac{\rho}{3M_4^2} \left( 1 + \frac{\rho}{2\lambda} \right)\, ,
\ee
where $\lambda$ is the tension of the brane, which is related to the five dimensional Planck mass as $\lambda=6M_5^6/M_4^2$. This equation says that at high energy densities the expansion of the universe was much faster than at later times, and went as $T^4$ instead of $T^2$, together with the unknown parameter $M_5$: the smaller $M_5$ the faster the expansion.

This five dimensional mass scale is constrained, in the ``RSII'' model, to be bigger than about $10^5$ TeV from measurements of the gravitational inverse square law in sub--mm range \cite{rs}. However such a bound might be inapplicable if the RSII model were only the low energy limit of a more fundamental theory \cite{no-bound}. In this case BBN only requires $M_5 \gtrsim 10$ TeV \cite{walker}.

At this point it is convenient to define a ``transition'' temperature $T_*$ from standard cosmology to brane one, which can be extracted from $\rho=2\lambda$ \cite{osamu}
\be\label{trT}
T_*^2 = 
\left(\frac{360}{\pi^2 \, g_*}\right)^{1/2}\, \frac{M_5^3}{M_4}\, ,
\ee
where $g_* = g_*(T)$ counts the relativistic degrees of freedom at a given temperature $T$. If the dominant component of the universe is not radiation then this ``temperature'' approximately means the fourth root of the energy density, and parametrises the epoch at which the transition occurs. In terms of this new quantity the Hubble parameter can be cast as
\be\label{newH}
H^2 = H_{4d}^2 \left[1 + \left( \frac{T}{T_*} \right)^4 \right]\, .
\ee
Here $H_{4d}$ stands for the standard four dimensional Hubble parameter.

This new expansion law needs to be taken into account when the amount of gravitino produced in the early universe is calculated, that is, this expression has to be plugged into (\ref{boltzdef}). Under the assumptions that $T_R\gg T_*$ and $T_*\gg T$, and that the extra dimension does not change the coupling of the gravitino zero mode to the matter residing on the brane, instead of (\ref{zeroYstd}) the abundance at the BBN is approximately given by \cite{osamu}
\be\label{zeroYbrane}
Y_{3/2}^0 = 3.5 \cdot 10^{-19}\, \left( 1 + \frac{\tilde m^2}{3 {m_{3/2}^0}^2} \right)\, \left( \frac{T_*}{\rm TeV} \right)\, .
\ee

The main point here is that the former constraints on $T_R$ need now to be imposed on $\sim2T_*$, and thus on the unknown five dimensional mass scale. This conclusion involved only the zeroth gravitino: sections \ref{branekk} and \ref{stdkk} are intended to extend the analysis to the full spectrum of KK modes.

\section{SUSY and extra dimensions}\label{xsusy}

Supersymmetry and supergravity in the context of extra dimensions has been investigated by several authors, primarily in connection with supersymmetry breaking by means of extra dimensional mechanisms \cite{xsusyA, xsusyB, xsusyC}. The main reason of interest on these models revolves around superstring theory, for it requires both supersymmetry and extra dimensions, although the path from such low--energy models and the full underlying string theory is far from being crystal clear. The cosmology of these models has not been studied yet, and, while it is expected that the well known main features of brane cosmology still hold, even relevant modifications could arise, primarily due to extra field in the bulk (the gravitino) and model--dependent orbifolding boundary conditions. This possibility is not explored further here, as the analysis presented in the forthcoming sections is readily extended to other cosmologies.

In order for this work to maintain its validity in a broad class of models, this analysis will be based mainly on two toy models, which reflect general features of supersymmetric extra dimension models in the literature. These toy models are a flat ADD--like one, in the spirit of \cite{xsusyA}, and a warped RS--like one, following \cite{xsusyB, xsusyC}. The extra dimension(s) is (are) compact.

\subsection{Flat bulk}\label{flatB}

In this model the bulk spacetime is flat and contains only gravitons and gravitinos. In considering the non--standard expansion epoch only the model with one extra dimension is analysed, since in this case the modified Friedman equation (\ref{modFried}) holds, whereas little is known for the general case.

The mass for each state can be expressed in two ways, depending upon the diagonalisability of the KK mass matrix\footnote{Henceforth the $n$--th KK gravitino mass will be just $m_n$.} \cite{xsusyA}
\be
m_n &=& m_0 + \frac{n}{R}\label{mADD}\\
m_n &=& \sqrt{ m_0^2+\left( \frac{n}{R} \right)^2 }\nonumber\, ,
\ee
where the first case holds if the KK mass matrix is diagonalised, while the other one does when it is not. Here $R$ is the size of the extra dimension, while $m_0$ is the zeroth mass, which can be either fixed by the extra dimensional parameters (this is the case if SUGRA is broken thanks to a mechanism which relies on the extra dimensions themselves), or not \cite{xsusyA}. Since there is no agreement on the way supergravity is broken, the zero mode mass will be taken as a free parameter, while for simplicity the mass matrix is assumed to be diagonalisable. That specified, the mass gap between two nearby states is given by
\be\label{gapADD}
\Delta m = \frac{1}{R} = \frac{2\pi M_5^3}{M_4^2} = 
\left( \frac{\pi^4 g_*}{90} \right)^{1/2}\, \frac{T_*^2}{M_4}\, .
\ee
This expression can be straightforwardly generalised to $N$ extra dimensions, except for the last equality.

Coming to the coupling constants, the situation is tricky and highly model dependent. Several distinct possibilities arise, as these couplings could be set by the extra dimension parameters, or be completely unrelated to them. This especially true for the $\pm 1/2$ helicity states \cite{xsusyA}, whereas $\pm 3/2$ ones should couple in the standard ($1/M_4$) way to brane--stuck MSSM matter. However, the goldstino states will reveal themselves to be not relevant in this study. Thus, the standard parametrisation for the cross section extracted from (\ref{zeroYstd}) is still valid, where of course the $n$--th KK gravitino mass $m_n$ has to be taken into account.

\subsection{Warped bulk}\label{warpB}

The second model to be dealt with is the warped one. Now a five dimensional cosmological constant resides in the bulk, which makes it an $AdS_5$ spacetime. Once again, the Friedman equation receives a high--energy correction as in (\ref{modFried}).

The mass spectrum is discrete, the KK modes masses being given by the following formula \cite{rs, rs-tower}
\be\label{mRS}
m_n = m_0 + k x_n e^{- \pi k R}\, ,
\ee
where $x_n$ is a solution of $J_1(x_n) = 0$ ($J_1$ is the BesselJ function of the first kind), $k$ is the $AdS_5$ curvature
\be
k = \frac{M_5^3}{M_4^2} \left(1 - e^{- 2 \pi k R}\right)\, ,
\ee
and $R$ parametrises the size of the extra dimension. The same hypothesis done for the flat model concerning the zero mode holds here as well. The mass gap reads
\be\label{gapRS}
\Delta m	&=& k e^{ -\pi k R} \left( x_n - x_{n-1} \right) \simeq 3 k e^{ -\pi k R} \\
					&=& \left (\frac{\pi^2 g_*}{40} \right)^{1/2} \frac{1 - e^{ -2\pi k R}}{e^{ \pi k R}}\, \frac{T_*^2}{M_4} \equiv \left (\frac{\pi^2 g_*}{40} \right)^{1/2} F(k R)\, \frac{T_*^2}{M_4}\, ,\nonumber
\ee
where $F(k R)$ is defined by the last equality.

The coupling constants in this case may be different for different modes. The reason for this is that the effective coupling on the brane is given by two factors, the actual coupling and the localisation of the wave function in the fifth dimension. Thus, a KK state peaked on the brane under inspection will interact strongly, while a state located in the other brane will be weakly interacting. The situation can be even more complicated if there is more than one tower of gravitinos, as it is likely the case since five dimensions require N=2 SUGRA at least.

A considerable simplification is made here by using again the standard cross section, that is, the KK gravitinos tower is taken to be localised in the far away brane. If instead the tower resided on ``our'' brane and the couplings reflected the ones for gravitons (Planck suppressed the zero mode, highly enhanced the KK states), KK gravitinos would thermalise, while the zero mode would not, and the resulting picture would reflect the one reviewed in section \ref{branesol}. However, in section \ref{TWstd} the simultaneous presence of two KK towers is investigated in more detail, in connection with the ``twisted'' model \cite{xsusyC}.

\section{KK gravitinos and standard cosmology}\label{stdkk}

Gravitinos are initially produced during a high temperature era. The total abundance for a given KK mode is computed by integrating (\ref{boltzdef}), where the $n$--th mode mass has to be taken into account. The upper limit for the integral is the highest temperature reached in the early universe for which the relativistic plasma was in thermal equilibrium; the lower limit is the temperature at which thermal production stops, which, for each mode, is approximately equal to its mass.

The abundance generated so far remains constant, except for some small jumps in the total entropy density, until it is time for these gravitinos to decay. The number density to entropy density ratio for the $n$--th gravitino mode at the BBN is
\be\label{nthYstd}
Y_{3/2}^n	&=& 1.9 \cdot 10^{-19}\, \left( 1 + \frac{\tilde m^2}{3 m_n^2} \right)\, \left( 1 - \frac{m_n}{T_R} \right) \left( \frac{T_R}{\rm TeV} \right)\, .
\ee

At this point both the zero mass and the mass gap are unspecified, hence, the calculation of the total amount of gravitinos could involve either an integral over the relevant range of masses, which is from $n = 0$ to $n = (T_R - m_0) / \Delta m \simeq T_R / \Delta m$, or a summatory over them. The result of the two operations (integral and summatory respectively) is,
\be
Y_{3/2}^{\rm tot}	&\simeq& 10^{-19}\, \left( \frac{T_R}{\rm TeV} \right) \{ \frac{T_R}{\Delta m} + \frac{2 \tilde m^2}{3 \Delta m^2} \frac{\Delta m}{m_0} \left( 1 + \frac{m_0}{T_R}\, \ln\frac{m_0}{m_0 + T_R} \right)\; \} \label{totYint1} \nonumber\\
									&\simeq& 10^{-19}\, \left( \frac{T_R}{\rm TeV} \right) \{ \frac{T_R}{\Delta m} + \frac{2 \tilde m^2}{3 \Delta m^2} \frac{\Delta m}{m_0}\; \} \label{totYint2} \\
\nonumber\\
Y_{3/2}^{\rm tot}	&\simeq& 10^{-19}\, \left( \frac{T_R}{\rm TeV} \right) \{ \frac{T_R}{\Delta m} + 1 + \frac{2 \tilde m^2}{3 \Delta m^2} \left\{ {\cal Y}_1[q] - {\cal Y}_1[q + x] \phantom{\frac{buh!}{aiuto!}} \right. \label{totYsum1} \nonumber\\
									&+& \frac{m_0}{T_R} \left. \left[ {\cal Y}_1[q] - {\cal Y}_1[q + x] + \frac{\Delta m}{m_0} \left( {\cal Y}_0[q] - {\cal Y}_0[q + x] \right) \right] \right\}\; \} \nonumber\\
									&\simeq& 10^{-19}\, \left( \frac{T_R}{\rm TeV} \right) \{ \frac{T_R}{\Delta m} + \frac{2 \tilde m^2}{3 \Delta m^2} {\cal Y}_1[q]\; \} \label{totYsum2}\, ,
\ee
where the function ${\cal Y}_a$ is the $a$--th logarithmic derivative the Gamma function, $q = m_0 / \Delta m$, and $x = (T_R + \Delta m) / \Delta m$. In order to obtain eqs.\ (\ref{totYint2}) and (\ref{totYsum2}), the condition $T_R \gtrsim m_0$ has been repeatedly used. As it can be easily seen these expressions are practically equivalent once the mass gap $\Delta m$ is smaller than the zero mode mass. Indeed, were this not the case then the integral would return a wrong answer: only (\ref{totYsum2}) would be reliable.

The lifetime of a heavy (${\cal O}(1)\, {\rm TeV}$) gravitino of mass $m_{3/2}$ is given, roughly speaking, by $M_4^2 / m_{3/2}^3$, while a light one is likely to be the LSP, since ${\cal R}$--parity conservation is assumed. Moreover, high KK--number gravitinos could decay into lighter ones, through processes such as $KK^n \rar KK^m + X$, where $m < n$ and $X$ is a Standard Model (SM) particle. This is connected with KK--number violation by the localisation of the process on the brane. Their amplitudes however can be either comparable or negligible with respect to the processes mentioned above (since also a transition $KK \rar LSP + X$ violates KK--number). Below the reasons why the inclusion of these processes should not significantly modify the results are outlined.

Thus, there are two possible scenarios, depending on the zero mode mass. If the zero mode is heavy it will decay producing a non--thermal abundance of LSP particles, as its KK tower will as well. However, there is a subtlety here. In the standard case, a gravitino which is heavy enough to decay when the temperature of the plasma is higher than the freeze--out temperature of the LSP pair annihilation process, it will not contribute to the non--thermal abundance for the LSP. This places an upper limit on the mass of the dangerous gravitinos (see the summary table in section \ref{stdsol}).

In standard cosmology the typical LSP freeze--out temperature is about 10 GeV, which corresponds to a time $t \sim (1 \; {\rm MeV}/T)^2 \sim 10^{-8}$ s. Since the gravitino lifetime is $\tau_{3/2} \sim 10^8 (100 \; {\rm GeV}/m_{3/2})^3$ s, gravitinos with masses larger than $m_{\rm MAX} \simeq 10^5 \; {\rm TeV}$ do not play any r\^ole. On the other hand, if KK gravitinos decay quickly into lighter ones, they could be dangerous as they will increase the non--thermal abundance of other KK states, in particular if there are many modes with masses in the range 100 GeV -- 10 TeV, where BBN constraints are quite strong.

If the zero mode is light the above considerations still hold approximately for each KK mode for which\footnote{The next--to--LSP (NLSP) mentioned here is not the first KK gravitino, but the lightest non--gravitino MSSM particle.} $m_n > m_{NLSP}$, for the heavy gravitinos decay into light SUSY particles which further decay into the LSP (which is the zero mode gravitino itself). In addition to the abundance therewith produced, one should thus take into account modes which survive, that is, those for which $m_n < m_{NLSP}$. Once more, if in addition we have transitions between KK modes, only these gravitinos whose lifetime is longer than the age of the universe $t_0$ can survive, while heavier modes would have decayed into them.

Summarising, with the aforementioned simplifications, the KK gravitino tower could be split into four ``bands'', keeping in mind that the lightest band may not exist if the zero mode is heavy enough. The first band consists of the modes for which $m_n < m_{NLSP}$: once produced they will remain as non--thermal relics. If there are direct transitions between KK modes only those for which $\tau_{3/2} > t_0$ contribute to the dark matter today, but their abundance is fed by the decays of heavier modes. It will be seen that these light modes are not going to be very relevant, though. The second band is that for which $m_{(N)LSP} < m_n < m_{MAX}$: these gravitinos end up as out--of--equilibrium LSP relics, whoever the LSP is. In the third band superheavy gravitinos live: they either decay into thermalised particles, and contribute nothing to non--thermal relics abundances today, or decay into lighter KK modes, increasing their abundances and tightening the constraints following from non--thermal gravitinos. The fourth band, which overlaps the second, and possibly also the first one, is the band for which gravitino decays affect BBN: this band may admit less freedom for the parameters of the models, and goes approximately from 100 GeV to 30 TeV.

\subsection{Flat extra dimension}\label{ADDstd}

A first rough constraint can be obtained by using the previously computed total amount of gravitinos. Moreover, since the relevant mass gaps are going to be around $\Delta m \gtrsim 100$ GeV and the temperatures of order a TeV or more, the gluino dependent part of the cross section can be safely neglected, since it is relevant for light gravitinos only.

The total amount of gravitinos\footnote{Notice that in this case eqs.\ (\ref{totYint2}) and (\ref{totYsum2}) are equivalent.} is then given by
\be\label{totYstdADD}
Y_{3/2}^{\rm tot} \simeq 3 \cdot 10^{-4} \frac{T_R^2}{M_4 \Delta m}\, .
\ee

A TeV zero mode gravitino forces the reheating temperature to be lower than order $10^2 \div 10^6$ TeV, see (\ref{example-T}) (the upper limit here reflects the fact that the gravitino could be much heavier that about 1 TeV). If, for example, $T_R^{(0)} \lesssim 10^5$ TeV is taken, where $(0)$ stands for zeroth gravitino constraint, then, for this to be the case, i.e.\ only one gravitino mode, the mass gap must be bigger than this temperature, or other modes will become available. This requirement gives a minimum mass gap, or a minimum size or Planck scale for the extra dimension, that is $M_5 \gtrsim 3 \cdot 10^{11}$ TeV. Were $M_5$ smaller, other KK states would have become available at $T_R^{(0)}$, and the limits on the reheating temperature would need to be reconsidered.

As an example consider $M_5 = 5 \cdot 10^9$ TeV, which in turn means a mass gap of about 100 GeV. If for every KK gravitino produced back then there is now a 100 GeV LSP particle, one can ask that this amount be smaller than $10^{-12}$, see table in section \ref{stdsol}. Plugging the mass gap into (\ref{totYstdADD}), the new limit on the reheating temperature reads $T_R \lesssim 10^3$ TeV.

For $B_h = 1$, the BBN constraint however is found to be stronger: since each of the 40 KK gravitinos in the mass range $1 \div 4$ TeV puts essentially the same bound $T_R \lesssim 10^3$ TeV \cite{abundance}, then $T_R$ is going to be 40 times smaller, that is $T_R \lesssim 25$ TeV. Gravitinos outside this mass span allow for much higher reheating temperatures, so that one can neglect them in this simple estimate. On the other hand, if $B_h = 10^{-3}$, the bound is weaker and comparable to the overclosure one. Indeed, looking at primordial abundances of $^6$Li and D, each one of the 40 KK gravitinos requires $T_R \lesssim 2 \cdot 10^5$ TeV, which becomes $T_R \lesssim 5\cdot10^3$ TeV once all the gravitinos are considered together.

Finally, it is noteworthy that all these reheating temperatures are below $T_*$, as required for consistency, for $M_5 > 10^7$ TeV.

Concluding, since the number of KK states available below a certain temperature grows very rapidly with the lowering of the five dimensional gravity scale, the corresponding allowed reheating temperature drops noticeably. If the mass gap is small, for instance around 1 keV, the reheating temperature would be tightly constrained around $T_R \simeq m_0$. This is no news since gravitons put similar upper bounds \cite{gravitons}. Furthermore, as for gravitons, stronger bounds would be deduced if there is more than one extra dimension and $M_5$ is not too big, because the number of KK states grows as $(T_R/\Delta m)^N$, where $N$ is the number of extra dimensions.

\subsection{Warped extra dimension}\label{RSstd}

The results of the previous section apply here as well, the only relevant difference being the expression for the mass gap, which is given by (\ref{gapRS}). Thus, for instance, the standard choice $k \simeq M_4$ and $k R \simeq 12$, gives a mass gap around 0.3 TeV. If this is the case, then the reheating temperature needs to be smaller than around $10^3$ TeV.

As a matter of fact, since these models involve SUSY and extra dimensions together there is no need for the extra dimension to solve the hierarchy problem. This implies more freedom in the choice of the values for the parameters, which may be significantly different from the example given above: (\ref{totYstdADD}) would still return the uppermost safe value for $T_R$.

It should be stressed here that these considerations are valid only if KK gravitino couplings are radically different from KK graviton ones. Indeed, only the zeroth graviton couples as $1/M_4$ to the brane, while KK gravitons interact with $1/M_5$ strength, since their wave functions are peaked on the SM brane. This means that KK gravitons are not able to give such restrictive constraints on the reheating temperature, while weakly coupled gravitinos are.

\subsection{Twisted extra dimension}\label{TWstd}

This subsection is devoted to some comments on the (seemingly realistic) possibility that if SUGRA is realised in an extra dimensional model then there will be more than one tower of KK gravitinos, and these towers may be not all localised on the same brane.

In view of \cite{xsusyC} one can build a model in which there are two towers of KK gravitinos, living on opposite branes. One tower will have strongly enhanced interaction strengths as KK gravitons have, and will most likely thermalise in the early universe, while the other tower will come with $1/M_4$ couplings. Hence, the scenario is almost the same as in the previous section, but the number of degrees of freedom for a given temperature is now different. In particular, if KK degrees of freedom are more numerous than MSSM ones, a different law for $Y_{3/2}$ will be found.

In order to illustrate this fact, one needs to specify the behaviour of the relativistic degrees of freedom with the temperature. In such a picture this quantity can be approximately expressed, for temperatures higher than the mass of the zero mode, as
\be\label{dofTW}
g_*(T) \simeq g_{\rm MSSM}(T) + \left( g_{3/2} + g_2 \right) \frac{T}{\Delta m}\, ,
\ee
where $g_{MSSM}(T)$ counts MSSM degrees of freedom and it is weakly dependent on the temperature, while the second factor accounts for the relativistic KK gravitons and gravitinos in equilibrium ($g_{3/2} = 4$, $g_2 = 5$). If the second term dominates, the total amount of {$1/M_4$ interacting KK gravitinos will be given by
\be\label{totYstdTW}
Y_{3/2}^{\rm tot} \simeq 10^{-4} \frac{T_R}{M_4}\, \left( \frac{T_R}{\Delta m} \right)^{3/2}\, .
\ee

The effect of equilibrium KK states is to produce more efficiently dangerous gravitinos. This can be quantified by choosing the values for the parameters as in the previous section, that is, $k \simeq M_4$ and $k R \simeq 12$. The corresponding reheating temperature is $T_R \lesssim 600$ TeV, whereas $2 \cdot 10^3$ TeV was found in sec.\ \ref{RSstd}. Note that the assumptions made in deriving (\ref{totYstdTW}) hold here, since there are more than $10^4$ KK states available, while the MSSM degrees of freedom are much less, and since with this choice of parameters KK gravitinos interact strongly. Furthermore, KK gravitons are present in the model of section \ref{RSstd} as well, and the estimates given there would need to be improved taking these states into account if necessary.

\section{KK gravitinos and brane cosmology}\label{branekk}

\subsection{Flat extra dimension}\label{ADDbrane}

The above discussion can be generalised allowing for an epoch of non--standard expansion, which would have taken place after $T_R$ but before BBN. The Friedman equation is given by (\ref{newH}). The $n$--th mode abundance is thus given by (again the gluino term in the cross section is neglected):
\be\label{nthYbrane}
Y_{3/2}^n	\simeq 10^{-19} \left( \frac{T_R}{\rm TeV} \right) \phantom{x}_2F_1[ \frac{1}{4} , \frac{1}{2} ; \frac{5}{4} ; -\left( \frac{T_R}{T_*} \right)^4] \simeq 3 \cdot 10^{-19} \left( \frac{T_*}{\rm TeV} \right)\, ,
\ee
where $F$ is the Gauss' Hypergeometric function, and the last step implies $T_R \gg T_*$. This equation basically means that gravitinos are mainly produced around $T_*$, regardless of $T_R$ as long as it is much bigger than $T_*$ itself. This is the result obtained for the zero mode in \cite{osamu}.

Since all the gravitinos with masses lighter than $T_*$ are produced in the amount predicted by (\ref{nthYbrane}), while the production of heavier ones is strongly suppressed, the total gravitino abundance will be given by
\be\label{totYbraneADD}
Y_{3/2}^{\rm tot} \simeq 10^{-3} \frac{T_*^2}{M_4 \Delta m} \simeq 10^{-4}\, ,
\ee
where the last equality follows from (\ref{gapADD}). It is straightforward to conclude that KK gravitinos and the non--standard expansion epoch are not compatible with each other, the only possible way out being that all of the KK masses lie outside the range for which gravitinos are constrained by overclosure or BBN, but this seems unrealistic since it would require fine tuning of the zeroth mass together with $T_*$.
This can be seen in another way: the available number of KK states, inversely proportional to $T_*^2$, grows faster than the amount which can be cut away by lowering $T_*$ itself. Thus, once a small $T_*$ is taken, as demanded by the zeroth gravitino bound, many KK states would become available below that temperature, which would require a further step downwards for $T_*$, which in turn implies even more KK states available, and so on. There is no value for $M_5$ for which a safe enough amount of gravitinos is produced. This is entirely due to the relation between $T_*$ and $\Delta m$.

For instance, had the transition temperature been chosen around $10^5$ TeV, as imposed by the zeroth gravitino constraint, the mass gap would have been around $3 \cdot 10^{-5}$ TeV, which means an enormous number (about $10^9$) of KK states available at that temperature. 

\subsection{Warped extra dimension}\label{RSbrane}

In this case one could hope that since the mass gap depends on two unknown parameters there will be some parameters space for which the conclusion of the previous section could be evaded. However, despite this fact, unless the gravitino zeroth mass and the temperature scales on the scene are finely tuned, there is still no way one can get rid of the too many KK gravitinos.

In the warped case the mass gap is given by (\ref{gapRS}), and the overall amount of gravitinos becomes
\be\label{totYbraneRS}
Y_{3/2}^{\rm tot} \simeq 10^{-3} \frac{T_*^2}{M_4 \Delta m} 
\simeq \frac{10^{-3}}{F(k R)} \, .
\ee

One can easily see that this case is not better than the previous one, since the function $1/F(k R)$ is always bigger than approximately 2.5. This means that $Y_{3/2}^{\rm tot} > 3 \cdot 10^{-3}$, which is of course too much. If the zeroth gravitino constraint is imposed one would find $\Delta m / {\rm TeV} \simeq 10^{-5} F(k R) \lesssim 3 \cdot 10^{-6}$: tons of KK states are available in this scenario as well.

One more comment is in order here. As in section \ref{RSstd}, KK gravitons, if in thermal equilibrium, could play an important r\^ole, namely they could enhance the abundance given by (\ref{totYbraneRS}). If that were the case then (\ref{totYstdTW}) would approximately give the right amount of KK gravitinos at the BBN, where $T_R$ needs to be replaced by $2 T_*$. However, after adding these extra states the result (\ref{totYbraneRS}) does not change much (recall it is an order of magnitude estimate).

To conclude, it appears very difficult to reconcile an epoch of non standard expansion and the presence of a KK tower of gravitinos, at least in these simplified toy models, unless one demands a fine tuning between the parameters of the model. It should be stressed once more here that if the coupling constants are drastically different these conclusions do not hold; in particular, were KK gravitinos strongly interacting (as KK gravitons in RS--like models) they would be part of the thermalised plasma, which could not provide any constraint on the extra dimension free parameters.

\section{More Gravitinos}\label{mech}

In this section other gravitino production mechanisms are very briefly discussed: supersymmetric particle decays and non--thermal generation by scalar fields or by time variable background metric of an expanding universe. 

\subsection{Thermal SUSY particle decay}\label{decay}

In addition to inelastic scattering, gravitinos can be produced in a hot plasma by decay of supersymmetric particles \cite{thermal}. The partial decay width is roughly
\be\label{pdw}
\Gamma_X \sim \frac{1}{48 \pi} \frac{m_X^5}{m_{3/2}^2 M_4^2} \, ,
\ee
where $m_X$ is the mass of the initial state. In standard cosmology, such a production mechanism is relevant only for light gravitinos, i.e.\ for masses smaller than about 100 keV, but gives rise to very strong bounds: if $m_{3/2}$ is in the range $1 \div 100$ keV, $T_R$ cannot exceed the mass of SUSY particles ($0.1 \div 1$ TeV), or the production rate is so large that the universe would be overclosed. Such a low reheating temperature is probably impossible to realise in many inflationary scenarios. 

In braneworld cosmology the picture is essentially unchanged if there is a light zero mode and a mass gap not smaller than $T_R$: the decay width is always given by (\ref{pdw}) and, if the universe temperature is high enough that supersymmetric particles are relativistic and in thermal equilibrium (that is, their number density is not Boltzmann suppressed), the mechanism is too efficient for $m_{3/2} \simeq 1 \div 100$ keV. The possibility that after inflation the universe expansion is non--standard, i.e.\ $T_* \lesssim 1$ TeV, is unrealisable, because that would imply $\Delta m \lesssim 10^{-6}$ keV (see eqs.\ (\ref{gapADD}) and (\ref{gapRS})), which further opens way to a huge number of KK states. So, KK gravitinos and non--standard expansion in the early universe continue to be not compatible with each other, unless the extra dimensional SUSY--breaking mechanism provides significantly different couplings.

\subsection{Non thermal decays}\label{nonth}
Perturbative and non--perturbative gravitino production by scalar fields in the early universe was considered for the first time in \cite{non-thermal}. Depending on the particular framework, the mechanism may be completely negligible or the dominant gravitino source. In any case, it does not influence the other production processes and hence cannot reconcile the existence of gravitationally interacting KK gravitinos with non--standard expansion after inflation. 

\subsection{Gravitational production}\label{grav}

Gravitational particle creation in braneworld cosmology has been very recently discussed in \cite{us}. The mechanism is quite interesting, because it allows for the generation of many very weakly interacting or sterile particles, which today may account partially, or even completely, for the cosmological dark matter. The final abundance depends only on the particle mass and, in order to be non--negligible, the universe would have had to undergo a period of braneworld regime. However, since this mechanism represents an additional source of gravitinos during the non--standard expansion epoch, it does not help. On the contrary, even more dangerous particles would be produced.

\section{Some considerations about KK gravitons}\label{gravitons}

Up to now the focus has been mainly on gravitinos. It is quite natural to wonder whether similar conclusions could be deduced by considering gravitons alone, whose better known properties furnish more reliable grounds for discussing BBN constraints.

In ADD--like models, KK graviton interactions are $1/M_4^2$ suppressed, so, the corresponding lifetime is basically the same of KK gravitinos of equal mass, as long as we are not dealing with gravitino--goldstino states. However, since the graviton zero mode is massless and gravitons are not supersymmetric particles (and thus they have not a corresponding graviton {\cal R}--parity), only those whose masses lie within $100 \; {\rm GeV} \div 30 \; {\rm TeV}$ are useful. Indeed, heavier gravitons provide essentially no bounds, because their decay can not affect BBN or produce stable and dangerous relic particles. Concerning lighter gravitinos, only fairly weaker constraints can be deduced from BBN and CMBR, because their decay could spoil BBN predictions and/or produce distortions of the CMBR spectrum. Nevertheless, KK gravitons in the mass range $100 \; {\rm GeV} \div 30 \; {\rm TeV}$ should suffice, that is, KK gravitons and non--standard expansion in ADD--like models are not compatible as well.

The situation is completely different in RS--like models, as here KK graviton wavefunctions are peaked on ``our'' brane, so they interact much more strongly. In this case KK gravitons could thermalise, and no relevant bounds would be obtained. Of course if some modes do not reach thermal equilibrium, they would be able to constrain the reheating (or transition) temperature, even though these limits are expected to be much more shallow than what has been obtained previously.

\section{Conclusion}\label{end}

We have considered the phenomenology of toy models where supergravity is realised and subsequently broken in an extra dimensional setup, and we studied the cosmology of KK gravitino states which arise in that case. The most relevant conclusion is that, unless a considerable fine tuning between masses and parameters of the extra dimensional model is required, it is not possible to allow for an epoch of non--standard expansion and, at the same time, avoid KK gravitino overproduction. This is true for both flat and warped extra dimensional models, as long as there is at least one weakly interacting tower of KK gravitinos.

We obtained constraints on the reheating temperature of the universe when the latter is smaller than the transition temperature to standard expansion, i.e.\ when after inflation the hot universe started out following the standard Friedman law, and no modified expansion epoch has ever taken place thereafter. As expected, the availability of many KK states for a given temperature puts bounds on the reheating temperature which are stronger than in the standard case, where only the zero mode is present. We computed such upper limits in several illustrative scenarios, including a warped model in which a tower of strongly interacting gravitinos in thermal equilibrium coexists with a gravitationally interacting one, as in the model proposed in \cite{xsusyC}.

As far as high (${\cal O}(100)$ GeV or more) temperatures are concerned, our results are relatively general, as they do not rely on $\pm 1/2$ states which are significantly model--dependent. In regard to light KK gravitinos, general predictions are rather difficult to be made, since there is not a complete model of supersymmetry in extra dimensions.

We further discussed why other production mechanisms are likely to be unimportant, and commented on similar bounds coming from the KK tower of gravitons. In this connection there are two important differences. First of all, KK gravitons, if weakly interacting, provide constraints only if they decay after BBN, that is, only for a given range of masses, whereas KK gravitinos would produce stable particles (LSP) and must be demanded to not exceed the observed amount of cosmological dark matter. Secondly, in warped models KK gravitons interact strongly and in the early universe they thermalise, while KK gravitinos, with the assumptions explained through the paper, should not. Thus, much tighter limits can be obtained by investigating KK gravitinos cosmology in warped models, as we have shown. We notice once more that the presence of weakly interacting KK gravitinos is crucial in our estimates.

Concluding, we believe that our calculations provide an useful tool for studying the cosmology of supersymmetric extra dimensions, as any such model would need to deal with the constraints presented here.

\ack

We wish to thank A.D. Dolgov for useful comments and encouragement. F.U. thanks H. Murayama for kind hospitality at UC Berkeley where the discussion which originated this work took place. F.U. is supported by INFN under grant n.10793/05.

\end{document}